\documentclass[a4paper,11pt]{article}
\usepackage{pos}
\usepackage{multirow}

\title{Heavy meson lightcone distribution amplitudes from Lattice QCD}

\author[a,b]{Xue-Ying Han}

\author[c,d]{Jun Hua}
\author[e]{Xiangdong Ji}

\author[a,b]{Cai-Dian L\"u} 

\author[f]{Andreas Sch\"afer}

\author[e]{Yushan Su}

\author*[g,h]{Wei Wang}
\emailAdd{wei.wang@sjtu.edu.cn}

\author[i,j]{Ji Xu}

\author[k,l,m,n]{Yibo Yang}

\author[o]{Jian-Hui Zhang}
\author[p]{Qi-An Zhang } 

\author[q]{Shuai Zhao}
 
\affiliation[a]{Institute of High Energy Physics, Chinese Academy of Sciences, Beijing 100049, China}
\affiliation[b]{School of Physical Sciences, University of Chinese Academy of Sciences, Beijing 100049, China}

\affiliation[c] {Key Laboratory of Atomic and Subatomic Structure and Quantum Control (MOE), Guangdong
Basic Research Center of Excellence for Structure and Fundamental Interactions of Matter,
Institute of Quantum Matter, South China Normal University, Guangzhou 510006, China }
\affiliation[d]{Guangdong-Hong Kong Joint Laboratory of Quantum Matter, Guangdong Provincial Key Laboratory of Nuclear Science, Southern Nuclear Science Computing Center, South China Normal University, Guangzhou 510006, China}

\affiliation[e]{Maryland Center for Fundamental Physics, Department of Physics, University of Maryland, 4296 Stadium Dr., College Park, MD 20742, USA}

\affiliation[f]{Institut f\"ur Theoretische Physik, University Regensburg
D-93040 Regensburg, Germany}

\affiliation[g]{Shanghai Key Laboratory for Particle Physics and Cosmology, Key Laboratory for Particle Astrophysics and Cosmology (MOE), School of Physics and Astronomy, Shanghai Jiao Tong University, Shanghai 200240, China}

\affiliation[h]{Southern Center for Nuclear-Science Theory (SCNT), Institute of Modern Physics, Chinese Academy of Sciences, Huizhou 516000, Guangdong Province, China}

\affiliation[i]{School of Nuclear Science and Technology, Lanzhou University, Lanzhou 730000, China}
\affiliation[j]{School of Physics and Microelectronics, Zhengzhou University, Zhengzhou, Henan 450001, China}

\affiliation[k]{School of Physical Sciences, University of Chinese Academy of Sciences, Beijing 100049, China}
\affiliation[l]{CAS Key Laboratory of Theoretical Physics, Institute of Theoretical Physics,
Chinese Academy of Sciences, Beijing 100190, China}
\affiliation[m]{School of Fundamental Physics and Mathematical Sciences,
Hangzhou Institute for Advanced Study, UCAS, Hangzhou 310024, China}
\affiliation[n]{International Centre for Theoretical Physics Asia-Pacific, Beijing/Hangzhou, China}

\affiliation[o]{School of Science and Engineering, The Chinese University of Hong Kong, Shenzhen 518172, China}

\affiliation[p]{School of Physics, Beihang University, Beijing 102206, China}
 \affiliation[q]{School of Science, Tianjin University, Tianjin 300072, China}

\abstract{Lightcone distribution amplitudes (LCDAs)  within the framework of heavy quark effective theory (HQET) play a crucial role in the theoretical  description of weak decays of heavy bottom mesons. However, the first-principle determination of HQET LCDAs faces significant theoretical challenges. In this presentation, we introduce a practical approach to address these obstacles. This makes sequential use of effective field theories.  Leveraging the newly-generated  lattice ensembles, we present a pioneering lattice calculation, offering new insights into LCDAs for heavy mesons. Additionally, we discuss the impact of these results on the heavy-to-light form factors and briefly give potential future directions in this field. }

\FullConference{The XVIth Quark Confinement and the Hadron Spectrum Conference (QCHSC24)\\
 19-24 August, 2024\\
 Cairns Convention Centre, Cairns, Queensland, Australia\\}

\begin{document}
\maketitle

\section{Importance of heavy meson LCDAs}

Heavy meson light-cone distribution amplitudes (LCDAs) provide essential insights into the distribution of momentum between a heavy quark and a light anti-quark  within a heavy meson~\cite{Grozin:1996pq}. These LCDAs are formulated within the framework of heavy quark effective theory (HQET) as:
\begin{align}
\label{eq:HQETLCDA}
 \langle 0|O_{v}^P(t) &|H(p_H)\rangle = i \tilde f_H   m_H n_+\cdot v  \int_0^\infty d\omega e^{i \omega t n_+\cdot v} \phi_+^B(\omega;\mu),
\end{align} 
where $\omega$ is the momentum carried by the light quark,  $\tilde f_H$ is the HQET decay constant, and the involved  operators are $
O_{v}^P(t) = \bar q_s(tn_+) {n}\!\!\!\slash_+\gamma_5W_c[tn_+,0]{h}_v(0). $
In this context, $h_v$  represents the effective operator for a heavy quark moving in the $v$ direction, and $W_c[tn_+,0]$ denotes a finite-distance Wilson line along the light-like vector $n_+$  ($n_+^2=0$). The   $q_s$ stands for a light quark field with soft momentum, and $t$ signifies the separation along the light-cone between $q_s$  and $h_v$.

Heavy meson LCDAs are essential for describing exclusive reactions such as $B$ meson decays, and  in particular, are valuable in the following scenarios. 
\begin{itemize}
\item In nonleptonic decays of $B$ meson such as $B^0/\overline B^0\to \pi^+\pi^-$, which are important to explore the CP violation, the QCD factorization   can be employed for calculating the involved matrix elements~\cite{Beneke:2000ry}. The leading power factorization formula is given as: 
\begin{eqnarray}
\label{fform}
&\langle\pi^+\pi^-|Q_i|\overline B^0(p)\rangle =
f^0_{B\to\pi}(m_\pi^2)\int^1_0 dx\, T^{I}_i(x;\mu)\Phi_\pi(x;\mu)  \nonumber\\
&~~  +\int d\omega dx dy T^{II}_i(\xi,x,y;\mu)\phi_+^B(\omega;\mu) \Phi_\pi(x;\mu) \Phi_\pi(y;\mu).
\end{eqnarray}
Here $Q_i$ are the four-quark operators and $T_i^{I,II}$ are hard-kernels that can be perturbatively calculated.  $\phi_+^B$ is the HQET LCDAs, and $\Phi_\pi(x;\mu)$ is the LCDAs for the light meson. 

\item In flavor-changing neutral current processes such as $B\to K^*\ell^+\ell^-$, which are valuable for exploring new physics beyond the standard model, the factorization formula at leading power is provided as
\begin{eqnarray}
\langle K^*_a \ell^+\ell^-|{\cal H}_{eff}|B\rangle = T_a^I(q^2) \xi_a(q^2) + \sum_{\pm} \int_0^\infty \frac{d\omega}{\omega} \phi_{\pm}^B(\omega;\mu) \int_0^1 du \Phi_{K^*}(u; \mu)T_{a,\pm}^{II} (\omega, u, q^2), 
\end{eqnarray}
with $a={||, \perp}$ denoting the polarization of the $K^*$ meson.  $ \xi_a(q^2)$ are the soft contributions to the heavy-to-light transition form factors.

\item In heavy-to-light transition form factors at large recoil, both hard-scattering contributions $\Xi _a$ and soft contributions $\xi _a$ are at leading power.  Using soft-collinear effective theory, one can factorize the   $\Xi _a$   as a convolution of a perturbatively-calculable jet function $J_a$ and LCDAs:
\begin{eqnarray}
\label{fform}
&\Xi _a(q^2)= \int d\omega du J_a(\omega, u, q^2; \mu) \phi_+^B(\omega; \mu) \Phi_{K^*}(u;\mu),
\end{eqnarray}

\item Soft contributions to the form factor  can not be factorized directly.  In phenomenological method such as lightcone sum rules (LCSRs),  $\xi_a$ can be written in terms of HQET LCDAs~\cite{Gao:2019lta}: 
\begin{eqnarray}
\label{fform}
&\xi_a(q^2)  = C_a \int^{\omega_s}_0 d\omega' {\rm exp}\left[-\frac{n\cdot p \omega' -m_{K^*}^2}{n\cdot p \omega_M} \right][\phi^B_{-,eff}(\omega'; \mu)+ \phi^B_{+,m}(\omega'; \mu)],
\end{eqnarray}
where unavoidable model dependence on the so-called Borel parameter $\omega_M$ is  introduced. $n\cdot p= (m_B^2-q^2+m_{K^*}^2)/m_B$.  Combinations of the LCDAs, namely $\phi^B_{-,eff}(\omega', \mu)$ and $\phi^B_{+,m}(\omega', \mu)$  are introduced in the above equation~\cite{Gao:2019lta}. 
\end{itemize}


\section{Difficulties in first-principle calculations}

Obtaining a reliable determination of the complete distribution of heavy meson LCDAs from first principles is highly challenging for several reasons. Firstly, these quantities are defined on the light cone, rendering them challenging to directly handle in first-principle calculations such as lattice QCD. Secondly, the formulation of heavy meson LCDAs involves the HQET field $h_v$, which further complicates direct simulations on the lattice. Moreover, the coexistence of light-cone separation and the HQET effective field leads to cusp divergences, as exemplified by a perturbative one-loop result for the relevant operators~\cite{Braun:2003wx}: 
\begin{align}
& O_v^{P, \mathrm{ren}}(t,\mu) = O_v^{P, {\rm bare}}(t)   + \frac{\alpha_sC_F}{4\pi} \bigg\{\left(\frac{4}{\hat \epsilon^2} + \frac{4}{\hat \epsilon}\ln (it\mu)\right) O_v^{P, {\rm bare}}(t)  \nonumber\\
&\;\;\;\; - \frac{4}{\hat \epsilon}\int_0^1 du \frac{u}{1-u}[O_v^{P, {\rm bare}}(ut)-O_v^{P, {\rm bare}}(t)  ] \bigg\},
\end{align}
with  $d=4-\epsilon$ and the standard notation $2/\hat \epsilon=2/\epsilon -\gamma_E +\ln (4\pi)$.  The term ${4}\ln (it\mu) / {\hat \epsilon}$  is derived from the expansion of ${4 (it\mu)^\epsilon}/{\hat \epsilon^2}$, which results from the intersection of UV divergence and cusp divergence. As pointed out in Ref.~~\cite{Korchemskaya:1992je},  the heavy quark behaves similar to a Wilson line along the $v$ direction. The coefficient for ultraviolet (UV) divergences between two Wilson lines is proportional to the angle between these two Wilson lines, as defined in Ref.~\cite{Korchemskaya:1992je}: 
\begin{eqnarray}
\cosh\theta = \frac{n_+\cdot v}{\sqrt{n_+^2}{\sqrt{v^2}}},
\end{eqnarray}
in which  $v^2=1$. Apparently the light-cone divergence, $n_+^2=0$, directly leads to  the cusp divergence.

As another manifestation of cusp divergence,  it can be observed that in the local limit $t \to 0$, the $\ln(it\mu)$ term diverges. Consequently, the Operator Product Expansion (OPE) for heavy meson LCDAs can not be applied, leading to uncertainties in defining the non-negative moments of heavy meson LCDAs. Therefore unlike light meson LCDAs, the conventional method of calculating these moments is also unsuccessful in this context.

\section{Proposal with  the Sequential effective theory} \label{sec:theory}

We will demonstrate that by sequentially applying two effective theory factorization formulas,   Large Momentum Effective Theory (LaMET)~\cite{Ji:2013dva,Ji:2020ect} and boosted Heavy Quark Effective Theory (bHQET), one can successively isolate the higher two scales $P^z$ and $m_H$  to derive the  HQET LCDAs.  This provides a realistic approach to calculate the HQET LCDAs~\cite{Han:2024min,Han:2024yun}.

To address the cusp divergence, one can avoid using the HQET field $h_v$
  and instead directly construct the LCDAs with the QCD heavy quark field. Meanwhile, it is crucial to ensure that the infrared behavior of the two quantities involved is identical, with their differences only manifesting at the perturbative scale of ${\cal O}(m_H)$~\cite{Han:2024min,Han:2024yun}. An expansion by region analysis of QCD LCDAs and HQET LCDAs can be utilized to verify this assertion~\cite{Beneke:2023nmj,Deng:2024dkd}. This process involves boosting the heavy meson and employing factorization to separate the hard-collinear and soft-collinear degrees of freedom. In addition, lattice simulations should refrain from using lightcone quantities, which can be achieved by employing an equal-time correlator with a large momentum $P^z$. Following the factorization of hard and collinear degrees of freedom and taking into account the ultraviolet distinctions, one can convert the equal-time correlator to QCD LCDAs within a low-energy effective theory. By doing so, we can successfully address the cusp divergence, along with the other two obstacles, by utilizing a quasi distribution amplitude that encompasses three scales with hierarchical ordering: $P^z \gg m_H \gg \Lambda_{\textrm{QCD}}$~\cite{Han:2024min}.

It is crucial to emphasize that exploiting quasi Distribution Amplitudes (quasi-DAs) has the potential to overcome all three challenges encountered in calculating HQET LCDAs, but the process must be carefully devised. Specifically, in the large momentum limit, the quasi-DA encompasses three distinct scales: the large momentum $P^z$, the heavy hadron mass $m_H$ (which at leading power is equivalent to the heavy quark mass $m_Q$), and the hadronic scale $\Lambda_{\textrm{QCD}}$. It is important to note that only when $P^z \gg m_H \gg \Lambda_{\textrm{QCD}}$, the first two energy scales fall within the perturbative regime and can be successively integrated out. In the first step, the scale $P^z$  can be integrated out, facilitating the matching of the quasi-DA to the LCDAs defined in QCD. This matching is ensured by separating hard modes with offshellness $\left(P^{z}\right)^2$
 from collinear modes. This methodology aligns with the treatment of parton distribution functions (PDFs) and light meson LCDAs in  LaMET,  with the distinction that collinear modes in this context encompass both hard-collinear and soft-collinear modes. Subsequently, once the QCD LCDA is derived, the second step involves integrating out the $m_H$ scale and matching the resultant quantity to the LCDA defined in HQET, specifically within the framework of boosted HQET. Hard-collinear modes with off-shellness of $m_Q^2$ are absorbed by the jet function, while the low-energy contributions are incorporated into the HQET LCDAs~\cite{Ishaq:2019dst,Beneke:2023nmj}.


\section{Lattice QCD simulations and Numerical Results}

Currently, conducting a dynamical simulation of a fast-moving bottom meson is not feasible on the lattice, but  HQET LCDAs are independent of heavy quark mass $m_Q$. This is known as heavy flavor symmetry.  Hence, in this approach, there is freedom to select $m_Q$ as the charm quark mass $m_c$, the bottom quark mass $m_b$, or any other appropriate value, without affecting the resulting HQET LCDAs. Discrepancies in the obtained outcomes  may signal the influence of power corrections. For the upcoming simulation, we will opt for $m_Q=m_c$.


Numerical simulation in this study utilizes gauge configurations generated by the CLQCD collaboration, employing $N_f=2 + 1$ flavors of stout smeared clover fermions and the Symanzik gauge action \cite{Hu:2023jet,Zhang:2021oja}.   For this investigation, we focus on a specific ensemble, denoted as “H48P32” in Ref. \cite{Hu:2023jet}, characterized by the finest lattice spacing of $a=0.05187$ fm and a volume of $48^3\times144$. The sea quark masses are set to correspond to $m_{\pi}=317.2$ MeV and $m_K=536.1$ MeV. Moreover, we determine the charm quark mass by adjusting it to yield a $J/\psi$ mass consistent with its physical value of $m_{J/\psi}=3.96900(6)$ GeV within an accuracy of $0.3\%$. The heavy $D$ meson mass on this ensemble is established as $m_D=1.917(5)$ GeV.
Out of a total of 549 gauge configurations, we analyze 8784 measurements of the bare matrix element of the heavy $D$ meson with boosted momentum $P^z=\{0,~6,~7,~8\}\times2\pi/(La)\simeq\{0,~2.99,~3.49,~3.98\}$ GeV, respectively.   Details of the simulation and the results can be found in Ref.~\cite{Han:2024yun}.

\begin{figure}[!t]
  \centering
  \includegraphics[scale=0.7]{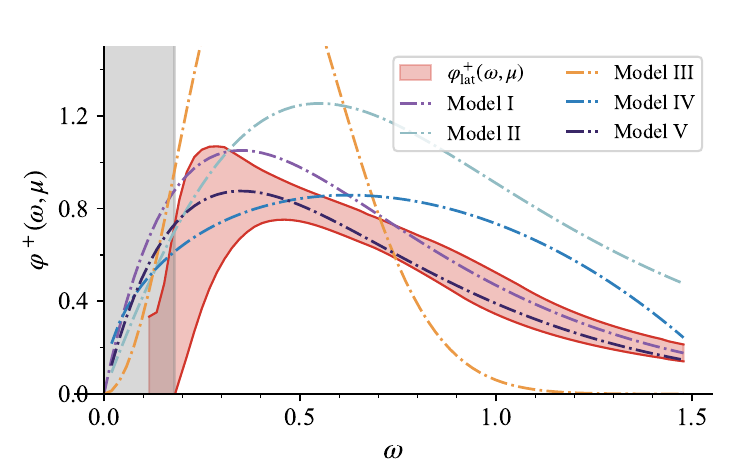}
  \caption{Comparison of the numerical results of $\varphi^{+}\left(\omega, \mu\right)$ from lattice QCD with several phenomenological models (see Ref.~\cite{Han:2024yun} for details). The renormalization scale $\mu$ of our result is chosen to be $m_D$.   }
  \label{fig:comparetopheno}
  \end{figure}

Based on this setup, we simulate the quasi DAs, and proceed with the renormalization of the bare matrix elements (MEs) in the hybrid-ratio scheme.  To reconstruct the complete distributions in coordinate space,  we make the following assumption, drawing inspiration from the asymptotic behavior in the long-tail region $e^{-\delta m z}c_1/|\lambda|^{d_1}$ with $\lambda= z\cdot P^z$.  With this form, we extrapolate the available results to the large $\lambda$ region. This can be used for Frourier transformation to momentum space. The quasi DAs can then be matched to QCD LCDAs through a LaMET factorization, which can be further matched HQET LCDAs. Our final results for the HQET LCDAs, after the renormalization, the $\lambda$ extrapolation and perturbative matching,  are given in Fig.~\ref{fig:comparetopheno}, where phenomenological pametrizations are also shown for a comparison.

The first inverse moment $\lambda_B^{-1}$ and the inverse-logarithmic moments $\sigma_B^{(n)}$  are crucial elements in QCD factorization theorems and lightcone sum rule investigations within heavy flavor physics. To extract these parameters from our developed HQET LCDA,   we utilize a model-independent parametrization formula to extrapolate our outcomes towards the vicinity of $\omega=0$ with the form: 
\begin{align}
\varphi^{+}&\left(\omega, \mu\right)=\sum_{n=1}^{N} c_n\frac{\omega^n}{\omega_0^{n+1}} e^{-\omega / \omega_0}  = \frac{c_1\omega}{\omega_0^2} \left[ 1 + c_2'\frac{\omega}{\omega_0} + c_3'\left(\frac{\omega}{\omega_0}\right)^2 + \cdots \right] e^{-\omega / \omega_0},
\label{eq:HQETLCDAsmallomegaexpansion}
\end{align}
with $c_n'\equiv c_n/c_1$. One advantage of this parametrization is that for small $\omega\ll \omega_0$, the impact of higher moments diminishes. By expanding Eq.(\ref{eq:HQETLCDAsmallomegaexpansion}) up to the first ($N=1$), second ($N=2$), and third ($N=3$) order terms, we can determine the fitting parameters, and it is found that outcomes demonstrate consistency, indicating a favorable convergence of the expansion.  

\begin{figure}[htbp]
\centering
\includegraphics[scale=0.3]{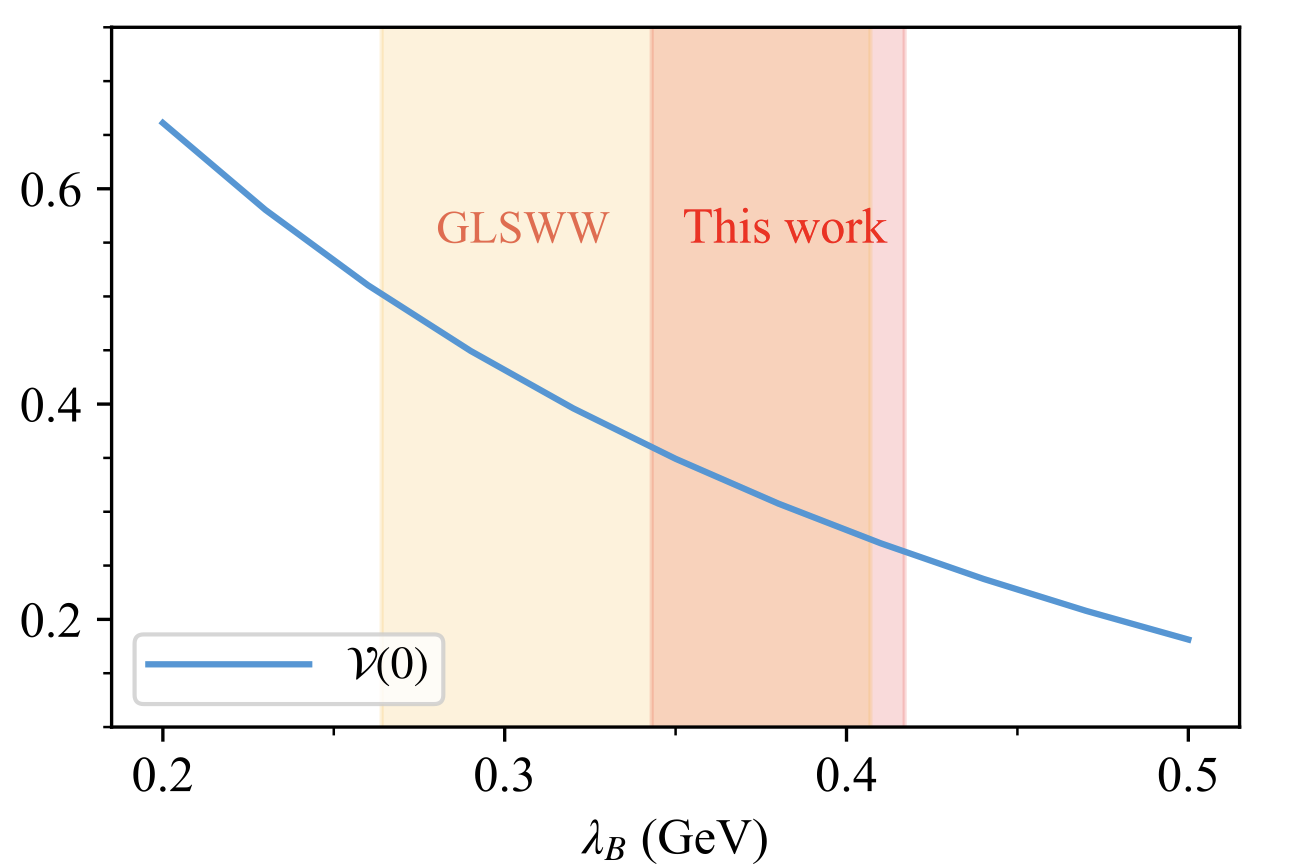}
\includegraphics[scale=0.3]{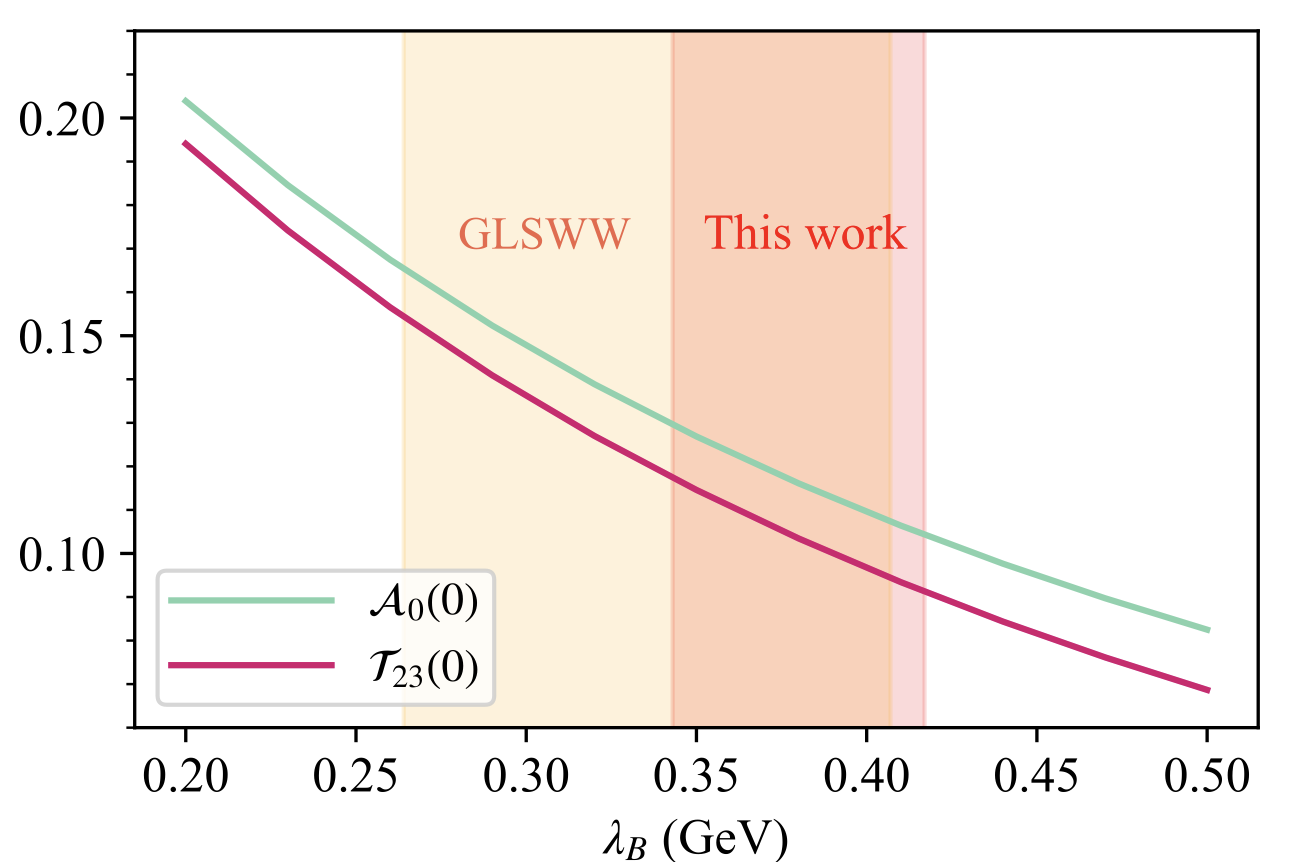}
\caption{The dependence of the $B\to K^*$ form factors on the inverse moment of heavy meson LCDAs.  The band labeled as ``GLSWW" represents the results obtained using the inverse moment directly from Ref.~\cite{Gao:2019lta}. On the other hand, the band labeled as ``This work" corresponds to the  results in this work. }
  \label{fig:B_Kstar_formfactors}
\end{figure}

Using these analytical results,  we present the dependence of the $B\to K^*$ form factors on the first inverse moment in Fig.~\ref{fig:B_Kstar_formfactors}. In alignment with the conventions of Ref.~\cite{Gao:2019lta}, we have adopted the results for $N=1$. Given that ${\cal V}$, ${\cal T}_1$, and ${\cal A}_1$ exhibit very similar numerical values, we have chosen to only display the results for ${\cal V}$ and ${\cal A}_0, {\cal T}_{23}$ at $q^2=0$.  From the results shown in Fig.~\ref{fig:B_Kstar_formfactors}, it is evident that the form factors calculated using our inverse moment approach align well with those presented in Ref.~\cite{Gao:2019lta}. It is important to note, however, that the systematic uncertainties have not been accounted for in this study. Examining the value of ${\cal V}^{B\to K^*}(0)$, it becomes apparent that a precise determination of $\lambda_B$ is crucial for achieving accurate predictions. Furthermore, our results derived from first-principles of QCD serve to eliminate the primary uncertainties associated with the model parametrizations of heavy meson LCDAs. This suggests the potential for a more refined analysis of form factors and, consequently, physical observables such as the $|V_{ub}|$
 .

\section{Summary and Prospects}

In this study, we present a novel framework for determining HQET LCDAs using a sequential effective theory approach.   We delve into the complexities of lattice QCD simulations, emphasizing the significance of analyzing dispersion relations and employing fitting techniques to extract meaningful outcomes. Our calculations showcase the quasi DAs, QCD LCDAs, and HQET LCDAs, providing valuable insights into the structure and behavior of heavy quark systems. Discussions center on the potential phenomenological implications of the findings, underscoring their relevance for future research in the realm of high-energy physics. We explore the potential phenomenological implications of the results, offering insights into how these findings could enhance our understanding of the dynamics of strong and weak interactions.  Thorough examinations of different systematic errors are underway.


\section*{Acknowledgement}

We thank the CLQCD collaborations for providing us the gauge configurations with dynamical fermions~\cite{Hu:2023jet}, which are generated on the HPC Cluster of ITP-CAS, the Southern Nuclear Science Computing Center(SNSC), the Siyuan-1 cluster supported by the Center for High Performance Computing at Shanghai Jiao Tong University and the Dongjiang Yuan Intelligent Computing Center. This work is supported in part by Natural Science Foundation of China under grant No. 12125503, 12335003, 12375069, 12105247, 12275277, 12293060, 12293062, 12047503, 12435004, 12475098, 12375080, 11975051, 12205106, and by the National Key Research and Development Program of China (2023YFA1606000).  A.S., W.W, Y.Y and J.H.Z are supported by a NSFC-DFG joint grant under grant No. 12061131006 and SCHA 458/22. Y.Y is supposed by the Strategic Priority Research Program of Chinese Academy of Sciences, Grant No.\ XDB34030303 and YSBR-101.  J.H.Z is supported by CUHK-Shenzhen under grant No. UDF01002851. Q.A.Z is supported by the Fundamental Research Funds for the Central Universities. J.H is supported by Guang-dong Major Project of Basic and Applied Basic Research No. 2020B0301030008.
The computations in this paper were run on the Siyuan-1 cluster supported by the Center for High Performance Computing at Shanghai Jiao Tong University, and Advanced Computing East China Sub-center.  This work was partially supported by SJTU Kunpeng \& Ascend Center of Excellence.

\end{document}